\documentclass[showpacs,amsmath,amssymb,twocolumn,pra,superscriptaddress]{revtex4-2}

\usepackage{graphicx}
\usepackage{dcolumn}
\usepackage{bm}

\usepackage{color} 
\usepackage[dvipsnames, svgnames, x11names]{xcolor} 
\usepackage{enumerate} 

\begin{document}

\preprint{APS/123-QED}

\title{Multipartite steering verification with imprecise measurements}

\author{Zeyang Lu}
\affiliation{Key Laboratory of Smart Manufacturing in Energy Chemical Process, Ministry of Education, East China University of Science and Technology, Shanghai, China}

\author{Chan Li}
\affiliation{Key Laboratory of Smart Manufacturing in Energy Chemical Process, Ministry of Education, East China University of Science and Technology, Shanghai, China}

\author{Gang Wang}
\affiliation{Key Laboratory of Smart Manufacturing in Energy Chemical Process, Ministry of Education, East China University of Science and Technology, Shanghai, China}

\author{Zhu Cao}
\email{caozhu@tongji.edu.cn}
\affiliation{College of Electronics and Information Engineering, Tongji University, Shanghai, China}
\affiliation{Shanghai Research Institute for Intelligent Autonomous Systems, Tongji University, Shanghai, China}
\affiliation{Shanghai Institute of Intelligent Science and Technology, Tongji University, Shanghai, China}

\begin{abstract}
Quantum steering is a fundamental quantum correlation that plays a pivotal role in quantum technologies, but its verification crucially relies on precise measurements---an assumption often undermined by practical imperfections. Here, we investigate multipartite steering verification under imprecise measurements and develop a quantitative method that effectively eliminates false positives induced by measurement imprecision. A comparison with a device-independent approach demonstrates that our method accurately delineates the scope of valid verification. In a special case, our method also enables the verification of multipartite entanglement under nonideal conditions. These results substantially enhance the robustness of multipartite steering and entanglement verification against measurement imprecision, thereby promoting their applicability in realistic quantum technologies.
\end{abstract}

\maketitle

\section{Introduction}\label{sec1}
Quantum steering, originally introduced by Schrödinger in 1935 as a fundamental yet intriguing nonlocal phenomenon, describes the ability of one party to remotely influence the quantum state of the other party through local measurements \cite{schrodinger1935discussion, uola2020quantum}. 
Distinct from entanglement \cite{horodecki2009quantum} and Bell nonlocality \cite{brunner2014bell}, steering occupies an intermediate position within the hierarchy of quantum correlations \cite{werner1989quantum, wiseman2007steering}.
Specifically, in bipartite steering scenarios, the measurement apparatus of one observer (the untrusted party) is treated as an unknown black box, while the measurement settings of the other observer (the trusted party) are assumed to be precisely characterized. 

In practice, however, achieving such precise control is often hampered by instrumental limitations and environmental noise. 
Even minor deviations from ideal measurement settings may lead to significant false positives, thereby undermining the reliability of quantum correlation tests involving trusted parties \cite{seevinck2007local, rosset2012imperfect}. 
For instance, in the context of entanglement witnesses, it has been shown that imprecise observables shift the corresponding bounds, which may cause separable states to be falsely identified as entangled if the imprecision is not properly taken into account \cite{morelli2022entanglement}.
A straightforward solution is to employ a device-independent approach (e.g., Bell nonlocality), in which all parties’ measurements are treated as black boxes \cite{bancal2011device, moroder2013device}. 
Although this method eliminates false positives, it severely restricts the scope of verifiability.
Recent studies have instead quantified the influence of measurement imprecision on verifying quantum correlations (e.g., entanglement and steering) in bipartite cases, thereby pinpointing the range of robust verification \cite{morelli2022entanglement, sarkar2023distrustful, tavakoli2024quantum}.  
Extending these analyses to multipartite systems is both natural and crucial, given their richer correlation structures and potential for deeper insights into the classical-quantum boundary \cite{roy2005multipartite, doherty2005detecting, palazuelos2022genuine, he2013genuine, cavalcanti2015detection, riccardi2018multipartite, jones2021network}. 
Indeed, Ref. \cite{cao2024genuine} has already explored how measurement imprecision affects the detection of genuine multipartite entanglement.  
Nonetheless, the analogous impact on multipartite steering verification remains elusive.

\begin{figure}[htbp]
\centering
\includegraphics[scale=1.1]{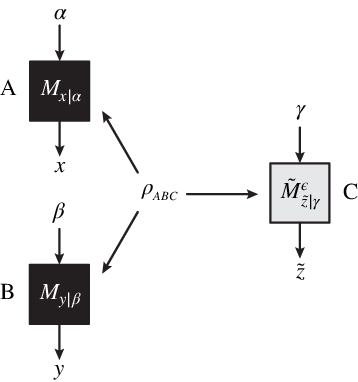}
\caption{
Schematic illustration of tripartite steering under imprecise measurements.
A tripartite state $\rho_{ABC}$ is distributed among Alice, Bob, and Charlie, with respective inputs $\alpha$, $\beta$, and $\gamma$ and corresponding outputs $x$, $y$, and $z$.
Alice and Bob are untrusted parties who perform uncharacterized measurements $M_{x|\alpha}$ and $M_{y|\beta}$, while Charlie is the trusted party.
In the presence of measurement imprecision $\epsilon$, Charlie’s ideal measurement $M_{z|\gamma}$ is replaced by the imprecise measurement $\tilde{M}_{\tilde{z}|\gamma}^{\epsilon}$, yielding the outcome $\tilde{z}$.
}
\label{fig_example}
\end{figure}

In this work, we address the verification of multipartite steering under imprecise measurements and introduce a quantitative approach to avoid false positives.
We focus on qubit systems with small but nonzero imprecision parameters, reflecting realistic experimental scenarios \cite{morelli2022entanglement, cao2024genuine, tavakoli2024quantum}.
Our method is based on a framework involving $N$ spatially separated parties, of which $T$ are trusted and $N - T$ are untrusted \cite{cavalcanti2011unified, he2011entanglement}.
This framework is flexible, as it allows for the adjustment of quantum correlation inequalities by varying $T$.
Specifically, when $T=N$, the inequalities detect multipartite entanglement; for $1 \leqslant  T < N$, they reveal multipartite steering; and when $T = 0$, they correspond to multipartite nonlocality.
As an example, Fig. \ref{fig_example} depicts the tripartite scenario with imprecise measurements for $N=3$ and $T=1$.

We derive a modified multipartite steering inequality that explicitly incorporates measurement imprecision.
In the special case $T = N$, our result also captures multipartite entanglement under imprecise measurements.
Our findings show that measurement imprecision increases the bounds of the multipartite inequalities when the trusted parties are involved, resulting in a gap between the modified bounds (considering imprecise measurements) and the original bounds (derived from ideal measurements). 
This gap expands significantly as the imprecision increases. 
If the original bounds are applied under conditions of measurement imprecision, observed statistics that lie within this gap may lead to false positives.
We illustrate this phenomenon with numerical simulations, which visually depict the region corresponding to these false positives.

Using the Greenberger-Horne-Zeilinger (GHZ) states \cite{pan2000experimental, bouwmeester1999observation, gottesman1999demonstrating}, we reveal that measurement imprecision affects the verification in markedly different ways depending on $N$.
Furthermore, comparison with the device-independent approach shows that our quantitative method effectively eliminates false-positive verifications and accurately delineates the scope of valid verification.
Therefore, our results significantly enhance the reliability of multipartite steering and entanglement verification in the presence of imprecise measurements, thereby advancing their applicability to practical quantum technologies such as quantum communication \cite{branciard2012one, gehring2015implementation, walk2016experimental, reid2013signifying, he2015secure}, randomness certification \cite{law2014quantum, passaro2015optimal, skrzypczyk2018maximal, li2024randomness}, and quantum computing \cite{li2015certifying, gheorghiu2017rigidity}.

The structure of the paper is summarized as follows.
Section \ref{sec2} introduces the general framework involving N spatially separated parties. 
In Sec. \ref{sec3}, we provide a quantitative characterization of measurement imprecision and derive a modified multipartite inequality that explicitly incorporates imprecision parameters.
Section \ref{sec4} investigates the impact of imprecise measurements from various perspectives through detailed numerical simulations.
Finally, Sec. \ref{sec5} concludes the study and discusses potential directions for future research.

\section{Framework} \label{sec2}
We consider a multipartite scenario involving $N$ spatially separated parties, where each party $k$ $(k=1,...,N)$ performs a measurement $X_k$ and obtains an outcome $x_k$.
Following Refs. \cite{cavalcanti2011unified, he2011entanglement}, we introduce the multipartite local hidden state (LHS) model, denoted as LHS($T, N$), where $T$ represents the number of trusted parties and $N-T$ denotes the number of untrusted parties. 
The LHS($T,N$) model is characterized by the following joint probability distribution:
\begin{equation}
    \begin{split}
        &p\left ( x_1, \dots , x_N | X_1, \dots , X_N\right ) \\ &= \int d\lambda p\left (\lambda \right )\displaystyle\prod_{k=1}^{T}p_Q\left ( x_k | X_k, \rho_{k,\lambda}\right )\displaystyle\prod_{k=T+1}^{N}p\left ( x_k | X_k, \lambda \right ),
    \end{split}
\end{equation}
where $\lambda$ is the hidden variable, $p(\lambda)$ is the probability distribution for $\lambda$, $p( x_k | X_k, \lambda )$ is known as the response function, and $\rho_{k,\lambda}$ is the hidden quantum state.
$p_Q( x_k | X_k, \rho_{k,\lambda} )$ refers to the probability distribution that follows the quantum mechanical measurement rule, which can be obtained by performing a positive operator-valued measure (POVM) $E_{x_k | X_k}$ on a quantum state $\rho_{k,\lambda}$, i.e., $p_Q\left ( x_k | X_k, \rho_{k,\lambda}\right ) = \mathrm{Tr}\left ( E_{x_k | X_k}\rho_{k,\lambda}\right )$.

A violation of the LHS($T, N$) model for $T = N$ serves as a criterion for verifying multipartite entanglement.
For $1 \leqslant  T < N$, such a violation is commonly referred to as multipartite steering.
It should be noted, however, that in this regime a violation does not necessarily imply steering in the strict sense; it may also originate from entanglement among trusted parties or from nonlocal correlations among untrusted parties.
When $T = 0$, it provides evidence of multipartite nonlocality.

To derive corresponding inequalities, we assume each party can measure either observable $X_k$ or $Y_k$, yielding the corresponding outcome $x_k$ or $y_k$, respectively.
We define the complex function $f_k^\pm = x_k \pm iy_k$ \cite{cavalcanti2007bell, he2011entanglement, cavalcanti2011unified}.
For qubit systems, this construction leads to the multipartite inequality
\begin{equation}\label{Pauli inequality}
    \begin{split}
        \left | \left \langle \displaystyle\prod_{k=1}^{N}f_k^{s_k}\right \rangle\right |^2 \leqslant  2^{N-T}  = B_0,
    \end{split}
\end{equation}
where $s_k\in \{-,+\}$, and the bound under ideal measurements is defined as $B_0$.
$X_k$ and $Y_k$ are chosen as mutually orthogonal Pauli observables.
One can vary the value of $T$ to obtain a family of multipartite inequalities tailored to the verification of different types of quantum correlations.

\section{Impact of imprecise measurements} \label{sec3}
We define measurements performed under precise control as target measurements $M_k$, which correspond to projective measurements. 
In contrast, measurements lacking precise control are referred to as laboratory measurements $\tilde{M}_k$, which are described by POVMs. 
For qubit systems, we quantify the imprecision of laboratory measurements by evaluating their proximity to the corresponding target measurements through 
\begin{equation}\label{measurement fidelity}
    \begin{split}
        \mathrm{Tr}\left ( M_k\tilde{M}_k\right ) \geqslant 1 - \epsilon_k ,
    \end{split}
\end{equation}
where $\epsilon_k\in [0,1]$ represents the imprecision parameter for the $k$th party. 
This fidelity condition is introduced solely as a quantitative measure of imprecision; it does not determine the structural form of the imprecise measurement.
In this work, we focus on cases where $\epsilon_k$ is small but nonzero \cite{morelli2022entanglement, tavakoli2024quantum, cao2024genuine}.

To incorporate measurement imprecision into the verification process, we model its effect on the observables in the original framework.
Specifically, the two observables of party $k$ are modified as
\begin{equation}
    \begin{split}
        \tilde{X}_k &= q_{k,x}X_k + \sqrt{1-q_{k,x}^2}X_k^\perp, \\ \tilde{Y}_k &= q_{k,y}Y_k + \sqrt{1-q_{k,y}^2}Y_k^\perp,
    \end{split}
\end{equation}
where $X_k^\perp$ and $Y_k^\perp$ denote observables orthogonal to $X_k$ and $Y_k$, respectively, satisfying $\mathrm{Tr}(X_kX_k^\perp)=0$ and $\mathrm{Tr}(Y_kY_k^\perp)=0$.
$q_{k,x}$ and $q_{k,y}$ are the alignment factors characterizing the imprecise observables.
From Eq. (\ref{measurement fidelity}), we obtain the inequality $\mathrm{Tr}\left ( X_k\tilde{X}_k\right ) \geqslant 2 - 4\epsilon_{k,x}$, which holds for all qubit observables.
It then follows that $1 - 2\epsilon_{k,x} \leqslant  q_{k,x}\leqslant  1$.
Similarly, we have $1 - 2\epsilon_{k,y} \leqslant  q_{k,y}\leqslant  1$.
For simplicity, we define $q_k = \min(q_{k,x}, q_{k,y})$, where $q_k\in[1-2\epsilon_k,1]$.
This choice ensures that the overall measurement deviation for the $k$th party is governed by the greater of the two imprecision parameters.

Under imprecise measurements, the complex function is modified to $\tilde{f}_k^\pm = \tilde{x}_k \pm i\tilde{y}_k$, where $\tilde{x}_k$ and $\tilde{y}_k$ correspond to the outcomes of the observables $\tilde{X}_k$ and $\tilde{Y}_k$, respectively. 
In this setting, the following inequality holds for any LHS($T, N$) model:
\begin{equation}
    \begin{split}
        \left | \left \langle \displaystyle\prod_{k=1}^{N}\tilde{f}_k^{s_k}\right \rangle\right |^2 \leqslant  \int d\lambda p(\lambda)\displaystyle\prod_{k=1}^{N}\left |  \left \langle \tilde{f}_k^{s_k} \right \rangle_\lambda\right |^2,
    \end{split}
\end{equation}
where $\langle \cdot \rangle_\lambda$ denotes the expectation value with respect to the hidden variable $\lambda$ and $\langle \tilde{f}_k^{\pm }  \rangle_\lambda  =  \langle \tilde{x}_k \rangle_\lambda \pm i \langle \tilde{y}_k \rangle_\lambda$.
For the untrusted parties, we have $|  \langle \tilde{f}_k^{\pm } \rangle_\lambda |^2 \leqslant  \langle \tilde{x}_k^2 \rangle_\lambda + \langle \tilde{y}_k^2 \rangle_\lambda$,
which follows from the non-negativity of the variance and holds for any hidden variable $\lambda$.
On the other hand, for the trusted parties, we derive the inequality
\begin{equation}
    \begin{split}
        \left |  \left \langle \tilde{f}_k^{\pm } \right \rangle_\lambda\right |^2 &= \left \langle \tilde{x}_k\right \rangle_\lambda^2 + \left \langle \tilde{y}_k\right \rangle_\lambda^2 \\ &\leqslant  \left ( 1 + 2q_k \sqrt{1-q_k^2}\right )\left ( \left \langle x_k^2\right \rangle_\lambda + \left \langle y_k^2\right \rangle_\lambda - C_k \right ),
    \end{split}
\end{equation}
where $C_k$ is a constant arising from the quantum uncertainty relation $\Delta_{\lambda}^2 \left ( x_k\right ) + \Delta_{\lambda}^2 \left ( y_k\right )\geqslant C_k$. 
To obtain the optimal bound for the inequality, we set $X^\perp = Y$ and $Y^\perp = X$. 
Without loss of generality, we assume $q_k \geqslant \sqrt{1 - q_k^2}$, which implies $\frac{\sqrt{2}}{2} \leqslant  q_k \leqslant  1$.
It follows naturally that $\epsilon_k$ is constrained within the range $0 \leqslant  \epsilon_k \leqslant  (2-\sqrt{2})/4$.
For qubit systems, it is convenient to select the Pauli matrices $\sigma_k^x$ and $\sigma_k^y$ as the observables, such that $X_k = \sigma_k^x$ and $Y_k = \sigma_k^y$.
The constant $C_k$ is consequently fixed at $C_k = 1$ \cite{hofmann2003violation}.
By combining these relations, we obtain the multipartite inequality
\begin{equation}\label{different imprecision inequality}
    \begin{split}
        &\left | \left \langle \displaystyle\prod_{k=1}^{N}\tilde{f}_k^{s_k}\right \rangle\right |^2 \leqslant  2^{N-T} \displaystyle\prod_{k=1}^{T} \left ( 1 + 2q_k \sqrt{1-q_k^2}\right ) \\ &\leqslant   2^{N-T}\displaystyle\prod_{k=1}^{T}\left ( 1 + 4\sqrt{\epsilon_k\left ( 1-\epsilon_k\right )} - 8 \epsilon_k\sqrt{\epsilon_k\left ( 1-\epsilon_k\right )}\right ),
    \end{split}
\end{equation}
which remains valid when the measurement imprecision varies across different parties.
A detailed derivation is provided in the Appendix.
For a conservative and uniform bound, we set $\epsilon = \max_{1\leqslant  k\leqslant  N} \epsilon_k$, with $0 \leqslant  \epsilon \leqslant  (2-\sqrt{2})/4$.
In this case, the inequality takes the form
\begin{equation}\label{same imprecision inequality}
    \begin{split}
        &\left | \left \langle \displaystyle\prod_{k=1}^{N}\tilde{f}_k^{s_k}\right \rangle\right |^2 \leqslant  2^{N-T} \left ( 1 + 2q \sqrt{1-q^2}\right )^T \\ &\leqslant   2^{N-T}\left ( 1 + 4\sqrt{\epsilon \left ( 1-\epsilon \right )} - 8 \epsilon \sqrt{\epsilon \left ( 1-\epsilon \right )}\right )^T = B_{\epsilon},
    \end{split}
\end{equation}
where $q\in [1-2\epsilon, 1]$, and the modified bound which accounts for measurement imprecision is defined as $B_{\epsilon}$.

For the primary regime of interest, $1 \leqslant  T < N$, a violation of Eq. (\ref{same imprecision inequality}) indicates multipartite steering under imprecise measurements.
In the special case $T = N$, such a violation verifies multipartite entanglement under nonideal conditions.
Notably, the imprecision parameter $\epsilon$ does not affect the bound for the untrusted parties, consistent with the assumption that their measurement settings remain uncharacterized. 
In contrast, the bound applicable to the trusted parties is sensitive to variations in $\epsilon$, increasing as imprecision grows.
This behavior reflects the fact that the measurements of the trusted parties are initially assumed to be perfectly known; however, as $\epsilon$ increases, the certainty of these measurements diminishes.

For $T \ne 0$, measurement imprecision causes the modified bound in Eq. (\ref{same imprecision inequality}) to exceed its original counterpart in Eq. (\ref{Pauli inequality}) derived under ideal conditions, thereby introducing a gap.
Applying the ideal bound in the presence of imprecision may lead to false-positive verifications.
Therefore, the modified bound provides a rigorous and reliable verification of multipartite steering and entanglement under measurement imprecision, which constitutes the main result of this work.

Our framework, which is based on the LHS($T,N$)-type inequality of Ref. [22], also connects to other approaches for quantum correlation verification under imprecise measurements. 
In particular, when restricted to the case of $N=2$ parties with a single trusted party ($T=1$), our framework reproduces a condition for bipartite steering under imprecise measurements.
This is conceptually consistent with the results of Ref. \cite{tavakoli2024quantum}. 
However, Ref. \cite{tavakoli2024quantum} builds on the high-dimensional steering inequalities developed in Refs. \cite{marciniak2015unbounded, skrzypczyk2015loss}. 
Consequently, the explicit algebraic bounds in Ref. \cite{tavakoli2024quantum} differ from ours.
More broadly, in the bipartite case with two trusted parties ($N=2$, $T=2$), our setting is related to Ref. \cite{morelli2022entanglement}, which employs entanglement witnesses rather than an LHS($T,N$)-based criterion. 
Although both yield experimentally testable bounds, their algebraic structures and derivations are distinct.
In the multipartite case, when $N=T$, our inequality verifies entanglement with the weaker requirement that at least two parties share entanglement, whereas Ref. \cite{cao2024genuine} provides criteria for genuine multipartite entanglement under imperfect measurements, which demands entanglement across all subsystems. 
Thus, our framework and Ref. \cite{cao2024genuine} address different levels of multipartite correlations.

By applying a first-order approximation to Eq. (\ref{same imprecision inequality}), we obtain
\begin{equation}
    \begin{split}
        \left | \left \langle \displaystyle\prod_{k=1}^{N}\tilde{f}_k^{s_k}\right \rangle\right |^2 \lesssim 2^{N-T}\left ( 1 + 4\sqrt{\epsilon }\right )^T,
    \end{split}
\end{equation}
which offers a clearer view of the dependence of the inequality bound on the imprecision parameter $\epsilon$.
Specifically, the bound becomes increasingly pronounced as $\epsilon$ grows.

\section{Numerical simulations} \label{sec4}
We analyze and illustrate the effects of imprecise measurements from three distinct perspectives.
To ensure consistency, we standardize the imprecision parameters across all involved parties, denoting these parameters by $\epsilon$, as defined in Eq. (\ref{same imprecision inequality}).

\begin{figure}[htbp]
\centering
\includegraphics[scale=0.37]{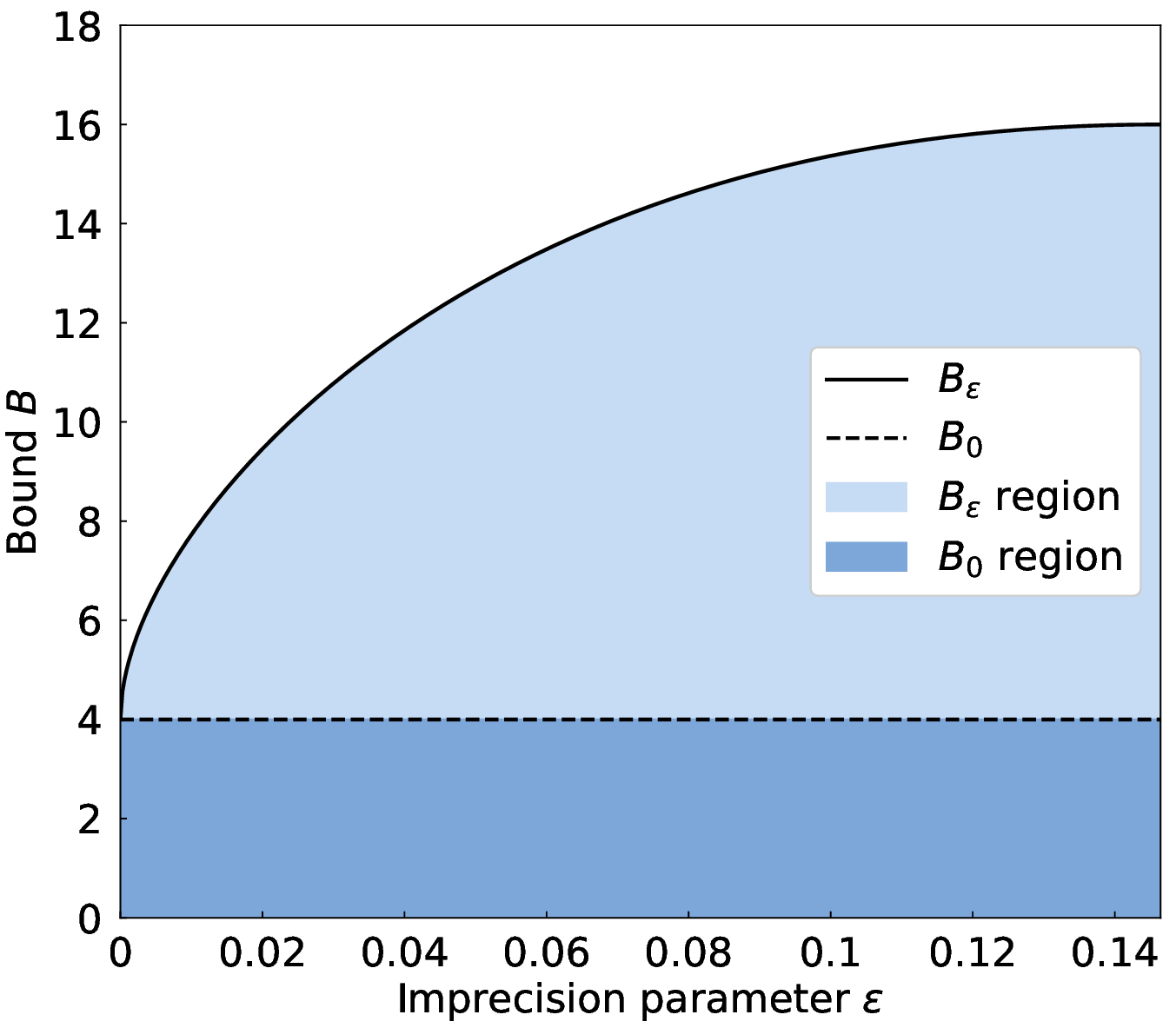}
\caption{
Simulation of inequality bounds $B$ as functions of the imprecision parameter $\epsilon$ for both ideal and imprecise measurement scenarios.
The total number of parties is set to $N=4$, with $T=2$ parties identified as trusted.
The solid line represents the modified bound accounting for measurement imprecision, denoted as $B_{\epsilon}$, while the dashed line indicates the original bound for ideal measurements, $B_0$. 
Shaded areas emphasize the region under each bound, with lighter shading corresponding to $B_{\epsilon}$ and darker shading corresponding to $B_0$.
}
\label{fig_false}
\end{figure}

We first address the issue of false positives induced by measurement imprecision. 
Figure \ref{fig_false} compares the original bound $B_0$ derived under ideal measurements with the modified bound $B_{\epsilon}$ which accounts for imprecision.
We consider a multipartite steering scenario involving $N=4$ parties, of which $T=2$ are trusted.
While $B_0$ remains constant, $B_{\epsilon}$ increases monotonically with $\epsilon$.
Verification of multipartite steering is deemed successful when the observed value on the left-hand side of the inequality exceeds the corresponding bound.
However, employing $B_0$ in the presence of measurement imprecision may lead to false positives when the observed value lies between $B_0$ and $B_{\epsilon}$.
The likelihood of such false positives increases with $\epsilon$.
Consequently, the modified bound $B_{\epsilon}$ is essential for reliable verification in nonideal measurement conditions.

We denote the left- and right-hand sides of Eq. (\ref{same imprecision inequality}) by $L$ and $R$, respectively.
The violation weight is then defined as $W = \sqrt{L / R}$, such that $W > 1$ indicates the presence of the corresponding multipartite quantum correlations.

\begin{figure}[htbp]
\centering
\includegraphics[scale=0.38]{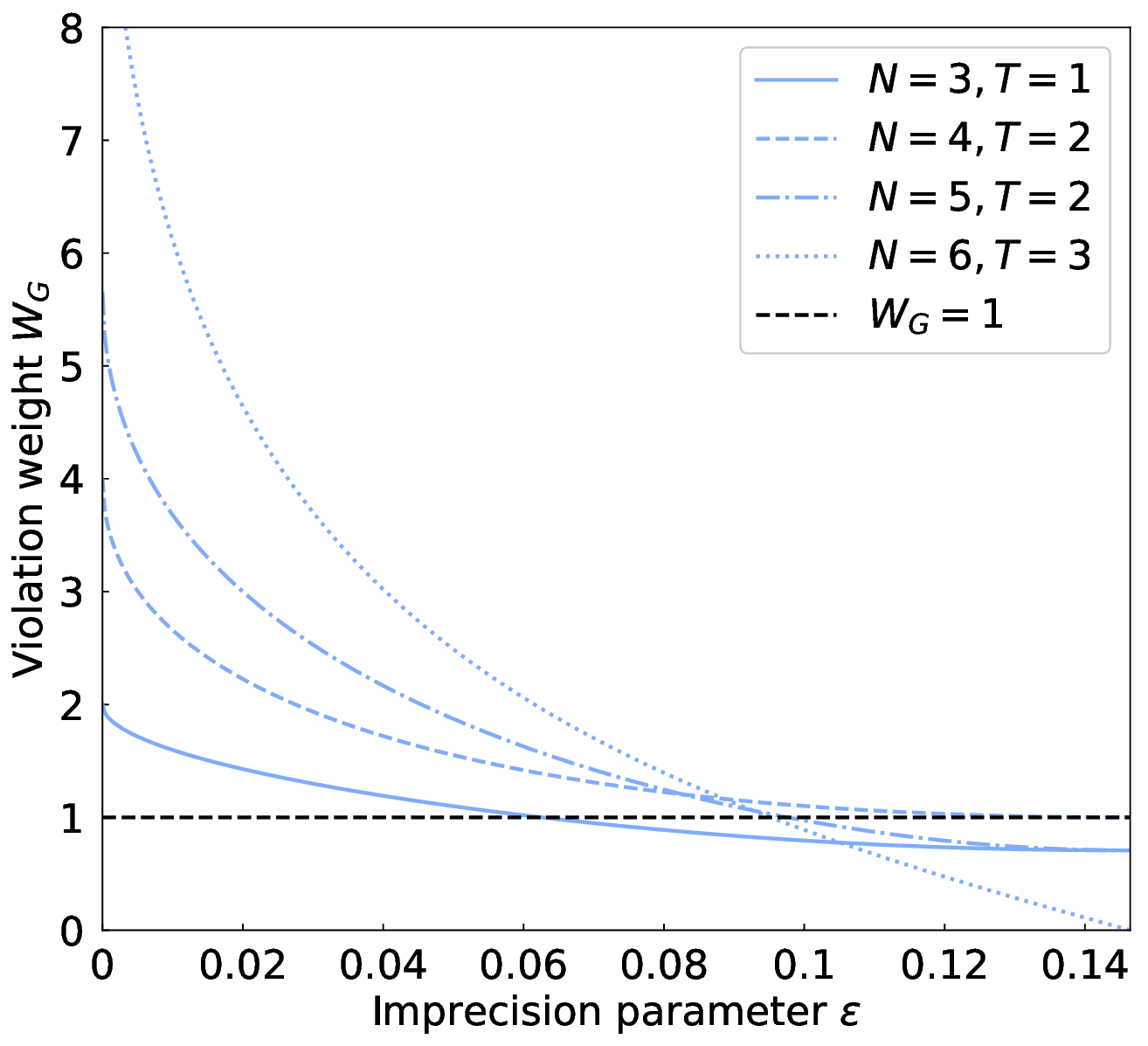}
\caption{
Simulation for the violation weights $W_G$ for multipartite steering as a function of the imprecision parameter $\epsilon$, for various values of the total number of parties $N$.
Four representative values of $N$ are shown, each distinguished by a different line type.
The number of trusted parties is set to $T = \lfloor N/2 \rfloor$.
The threshold $W_G=1$ is marked by the horizontal dotted line.
}
\label{fig_ste}
\end{figure}

We now examine the effect of measurement imprecision on multipartite steering verification for different numbers of parties $N$, as illustrated in Fig. \ref{fig_ste}. 
Our analysis employs the $N$-partite GHZ state, which is expressed as $ |\mathrm{GHZ}_N\rangle = \frac{1}{\sqrt{2}} (|0\rangle^{\otimes N} + |1\rangle^{\otimes N}) $.
For this state, the violation weight takes the form
\begin{equation}\label{W_G}
    \begin{split}
        W_G = 2^{\frac{N+T-2}{2}}\frac{\left | (1-2\epsilon)^N+i^N \left (2\sqrt{\epsilon(1-\epsilon)}\right )^N \right | }{\left ( 1-2\epsilon +2\sqrt{\epsilon(1-\epsilon)} \right )^T }, 
    \end{split}
\end{equation}
where $q = 1 - 2\epsilon$.
For clarity, we select four representative values of $N$ and set the number of trusted parties to $T = \lfloor N/2 \rfloor$. 
As $\epsilon$ increases, $W_G$ decreases overall, but the extent of this decrease depends on $N$. 
For small increments in $\epsilon$, $W_G$ remains above the threshold $W_G=1$ in all cases, with larger $N$ yielding higher values of $W_G$. 
However, as $\epsilon$ grows larger, distinct behaviors emerge: for $N=4$, $W_G$ gradually approaches the threshold; for $N=3$ and $5$, $W_G$ falls below 1; and for $N=6$, $W_G$ trends toward 0.
These differences arise from the presence of the factor $i^N$ in the calculations, which imparts a strong sensitivity to the specific value of $N$.
More generally, within the regime of interest $0 \leqslant  \epsilon \leqslant  (2-\sqrt{2})/4$, $W_G$ can be shown analytically to decrease monotonically as $\epsilon$ increases.
In particular, when $\epsilon = (2-\sqrt{2})/4$, denoted by $\epsilon_1$, we obtain
\begin{equation}
    \begin{split}
        W_G^{\epsilon_1}=\begin{cases}
0  & \text{ if } N = 4k - 2 , \\
\frac{\sqrt{2}}{2}  & \text{ if } N = 4k - 1 \ \mathrm{or}\ 4k + 1, \\
1  & \text{ if } N = 4k,
\end{cases}
\quad (k \in \mathbb{Z}^+).
    \end{split}
\end{equation}
At this point, the parameter $T$ does not affect the value of $W_G$. 
Moreover, these distinctions persist in the large-$N$ limit.

\begin{figure}[htbp]
\centering
\includegraphics[scale=0.38]{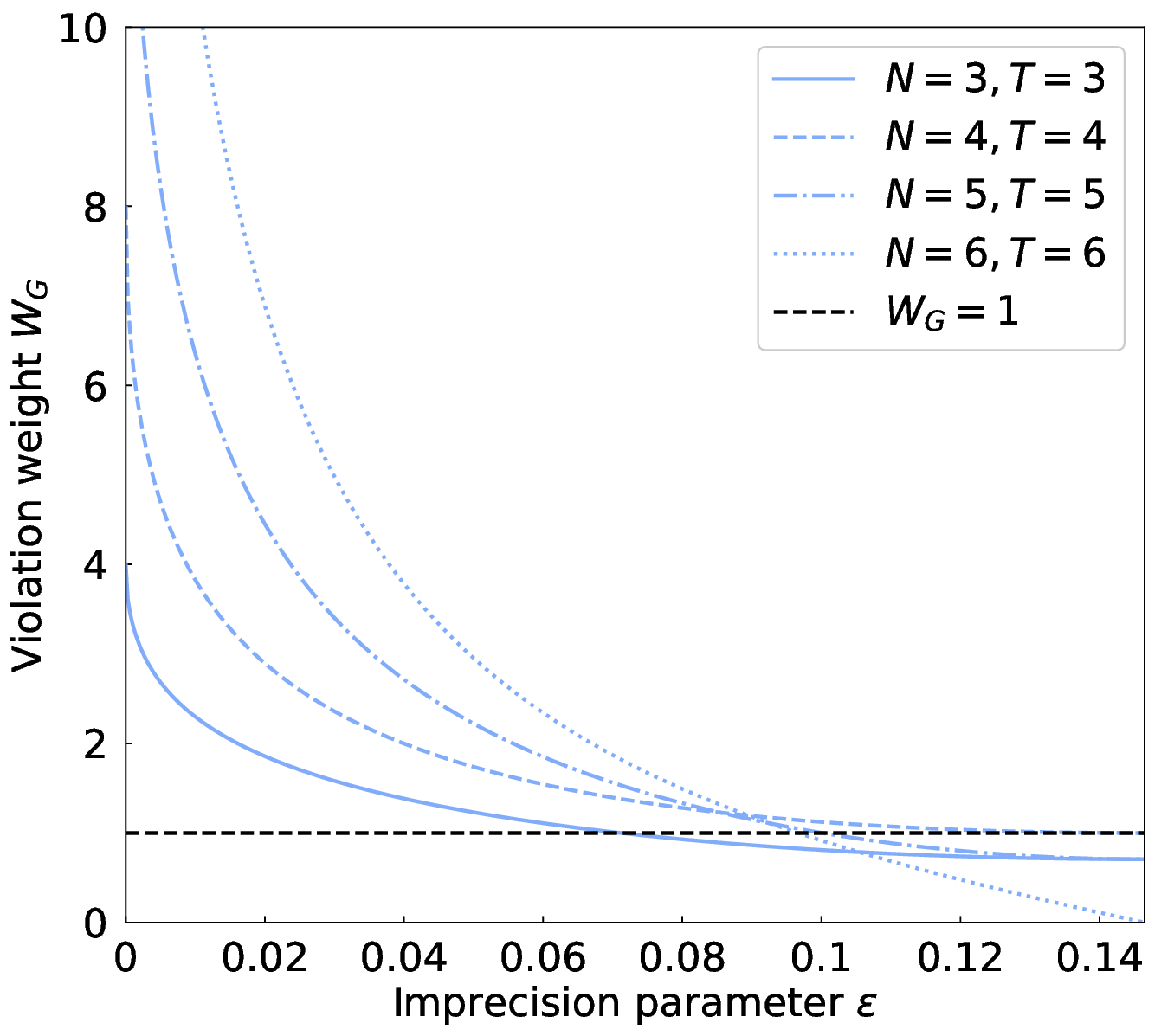}
\caption{
Simulation for the violation weights $W_G$ for multipartite entanglement as a function of the imprecision parameter $\epsilon$, for different values of the total number of parties $N$.
Each value of $N$ is represented by a distinct line style, with the number of trusted parties fixed at $T = N$.
The threshold $W_G=1$ is indicated by the horizontal dotted line.
}
\label{fig_ent}
\end{figure}

In Fig. \ref{fig_ent}, we similarly illustrate the influence of measurement imprecision on multipartite entanglement ($T = N$) for varying $N$.
The observed trends and their underlying causes align with those discussed for multipartite steering.
By comparing Fig. \ref{fig_ste} and Fig. \ref{fig_ent}, we observe that for a fixed $N$ and small $\epsilon$, a larger number of trusted parties $T$ leads to a larger value of the violation weight $W_G$. However, as $\epsilon$ increases, the influence of $T$ on the overall behavior of $W_G$ rapidly diminishes.

\begin{figure}[htbp]
\centering
\includegraphics[scale=0.37]{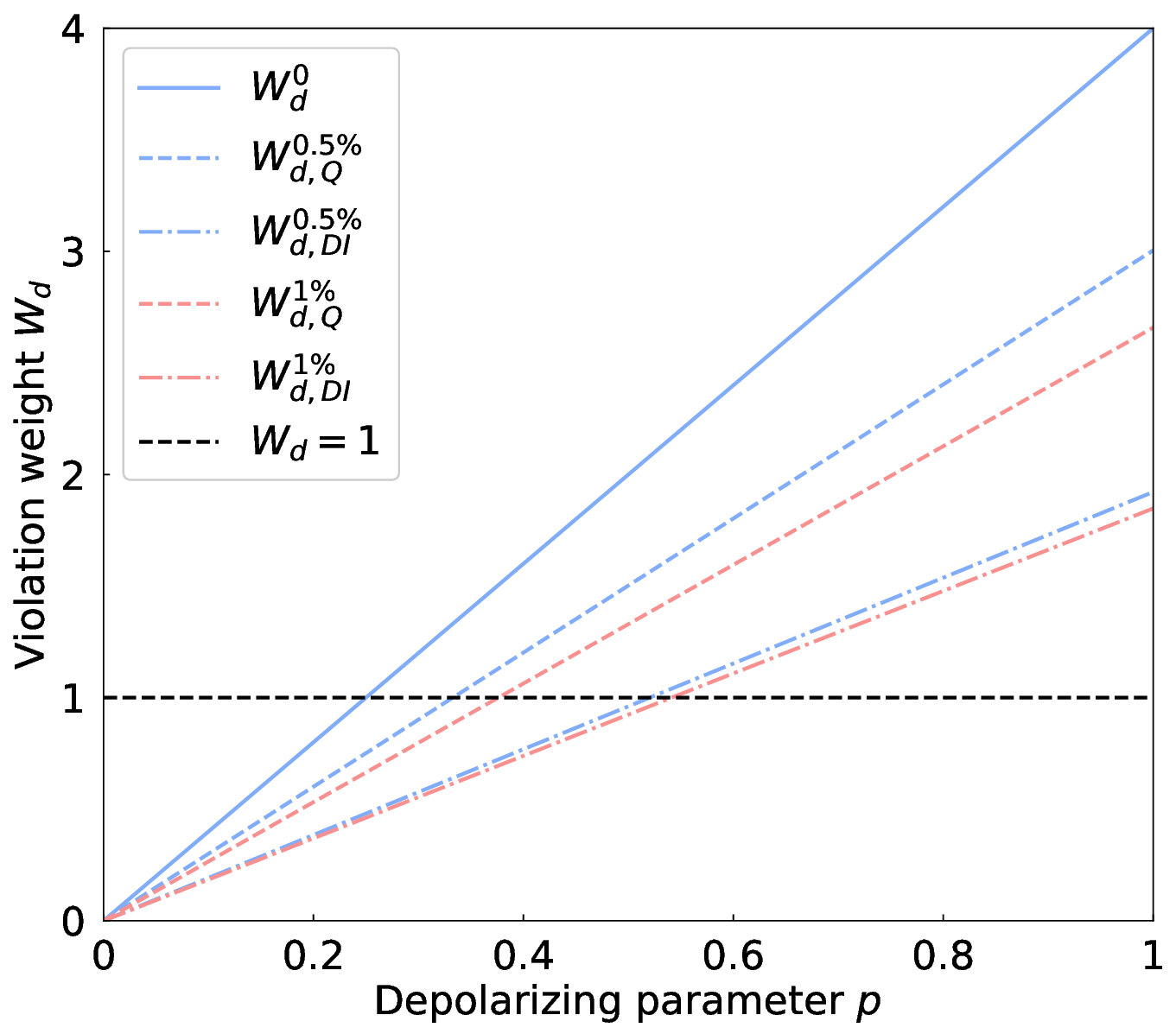}
\caption{
Simulation of the violation weight $W_d$ as a function of the depolarizing parameter $p$ under different methods.
The analysis is performed for the case where $N=4$ and $T=2$.
Three typical levels of imprecision parameter are considered: $\epsilon = 0$, $0.5\%$, and $1\%$.
Dashed lines represent the violation weight $W_{d,Q}$ obtained via the quantitative method, while dot-dashed lines indicate the violation weight $W_{d,DI}$ for the device-independent method.
For each method, a larger value of $\epsilon$ corresponds to a lower line position.
The horizontal dotted line marks the threshold $W_d=1$.
}
\label{fig_depolarized}
\end{figure}

To avoid false positives arising from imprecise measurements, one can employ a device-independent approach that treats the measurement devices of all parties as black boxes ($T=0$). 
While this approach obviates concerns about trusted devices, the resulting inequality bound is comparatively large, limiting the parameter range in which states can be verified. 
Figure \ref{fig_depolarized} compares the verification ranges of our quantitative approach with those of the device-independent approach under varying levels of measurement imprecision.
For concreteness, we analyze a depolarized GHZ state of the form
\begin{equation}
    \begin{split}
        \rho_d =p|\mathrm{GHZ}_N\rangle\langle \mathrm{GHZ}_N|+\left ( 1-p\right )\frac{\mathbb{I}}{2^N},
    \end{split}
\end{equation}
where $p$ denotes the depolarizing parameter with $0 \leqslant  p \leqslant  1$, and $\mathbb{I}$ is the $N$-partite identity operator. 
For this state, the violation weight is expressed as $W_d = pW_G$, with $W_G$ given in Eq. (\ref{W_G}).
The total number of parties is set to $N=4$, with $T=2$ trusted parties.
We consider three typical levels of imprecision parameter, $\epsilon=0$, $0.5\%$, and $1\%$, and denote the violation weights associated with the quantitative and device-independent methods as $W_{d,Q}$ and $W_{d,DI}$, respectively.
In the device-independent scenario, $R$ in $W_{d, DI}$ (i.e., the inequality bound) is independent of $\epsilon$ and therefore remains fixed. 
By contrast, $L$ in $W_{d, DI}$ (i.e., the observed value) is affected by $\epsilon$. 
As a result, the violation weight $W_{d,DI}=\sqrt{L/R}$ varies with $\epsilon$, which explains the behavior shown in Fig. \ref{fig_depolarized}.
In the ideal case $\epsilon=0$, the violation weight $W_d=1$ corresponds to $p = 0.25$.
For $\epsilon=0.5\%$ ($1\%$), $W_{d,Q}=1$ and $W_{d,DI}=1$ correspond to $p=0.33$ (0.38) and $p=0.52$ (0.54), respectively.
These results indicate that while the device-independent approach avoids the pitfalls of measurement imprecision, it significantly narrows the verifiable range of states. 
In contrast, by quantifying the effects of imprecise measurements, our approach not only ensures reliable verification but also identifies the range of verifiable states more accurately.

\section{Conclusion} \label{sec5}
In this study, we investigate how imprecise measurements resulting from practical constraints affect the verification of multipartite steering. 
We show that applying the bound derived from ideal measurements may result in false positives in the presence of imprecision.
To address this issue, we introduce a quantitative method that explicitly incorporates the measurement imprecision into the verification process, yielding a modified multipartite steering inequality. 
In particular, our method can also be applied to multipartite entanglement verification under realistic conditions.
We provide a visual and numerical analysis of false positives and show that our method effectively eliminates them.
Our simulations also reveal that the impact of imprecision on verification varies significantly with the number of parties $N$.
Furthermore, comparison with a device-independent method, which treats all parties' measurements as black boxes, confirms that our approach offers a more accurate and robust verification criterion.
These findings establish a practical and generalizable strategy for reliably verifying multipartite steering and entanglement in the presence of measurement imprecision.

The incorporation of imprecise observables into multipartite steering inequalities is not confined to the unified criterion of Ref. \cite{cavalcanti2011unified}.
In principle, the same procedure can be applied to other multipartite steering inequalities, such as those in Refs. \cite{he2013genuine}, \cite{cavalcanti2015detection}, and \cite{jones2021network}, with the differences arising mainly from their specific algebraic forms and underlying system assumptions.
This general applicability highlights a promising avenue for future exploration.

Moreover, the present framework may be extended to more intricate scenarios, including continuous-variable systems, higher-dimensional Hilbert spaces, and detection models with nonideal efficiencies.
It is worth noting that the detection inefficiency responsible for the detection loophole in device-independent tests is distinct from the measurement imprecision considered in this work. 
Detection inefficiency corresponds to missing outcomes and postselection on untrusted devices, whereas measurement imprecision refers to systematic misalignment of trusted observables.
Although both imperfections may occur simultaneously in practice, they affect the bounds in fundamentally different ways and therefore require separate modeling.

\begin{acknowledgments}
This work was supported by the Natural Science Foundation of Shanghai under Grant 25ZR1402098 and the startup fund from East China University of Science and Technology under Grant YH0142234.
\end{acknowledgments}

\begin{widetext}

\appendix
\renewcommand{\theequation}{A\arabic{equation}}
\setcounter{equation}{0}

\section*{Appendix: Proof of the impact of imprecise measurements}
We now derive the multipartite inequalities that quantify the effects of imprecise measurements, as discussed in the main text. 
Note that we focus on qubit systems.

Under imprecise measurements, the complex function is represented as $\tilde{f}_k^\pm = \tilde{x}_k \pm i\tilde{y}_k$. 
Here, $\tilde{x}_k$ and $\tilde{y}_k$ denote the measurement outcomes corresponding to the two-qubit observables $\tilde{X}_k$ and $\tilde{Y}_k$, respectively.
For any LHS($T, N$) model, the expectation value is given by
\begin{equation}
    \begin{split}
        \left \langle \displaystyle\prod_{k=1}^{N}\tilde{f}_k^{s_k}\right \rangle =\int d\lambda p(\lambda)\displaystyle\prod_{k=1}^{N}\left \langle \tilde{f}_k^{s_k}\right \rangle_\lambda,
    \end{split}
\end{equation}
where $s_k\in \{-,+\}$ and $\langle \cdot \rangle_\lambda$ denotes the expectation value with respect to the hidden variable $\lambda$.
It follows that $\langle \tilde{f}_k^{\pm }  \rangle_\lambda  =  \langle \tilde{x}_k \rangle_\lambda \pm i \langle \tilde{y}_k \rangle_\lambda$.
We then derive the inequality
\begin{equation}
    \begin{split}
        \left | \left \langle \displaystyle\prod_{k=1}^{N}\tilde{f}_k^{s_k}\right \rangle\right | \leqslant & \int d\lambda p(\lambda)\left | \left \langle \tilde{f}_1^{s_k}\right \rangle_{\lambda}\right |\left | \left \langle \tilde{f}_2^{s_k}\right \rangle_{\lambda}\right |\cdots \left | \left \langle \tilde{f}_N^{s_k}\right \rangle_{\lambda}\right | = \int d\lambda \sqrt{p(\lambda)}\sqrt{p(\lambda)}\displaystyle\prod_{k=1}^{N}\left | \left \langle \tilde{f}_k^{s_k} \right \rangle_{\lambda}\right | \leqslant  \left [ \int d\lambda \left ( \sqrt{p(\lambda)}\right )^2 \right ]^{\frac{1}{2}} \\& \times \left [ \int d\lambda \left ( \sqrt{p(\lambda)}\displaystyle\prod_{k=1}^{N}\left | \left \langle \tilde{f}_k^{s_k} \right \rangle_{\lambda}\right |\right )^2\right ]^{\frac{1}{2}} = \left ( \int d\lambda p(\lambda) \displaystyle\prod_{k=1}^{N}\left | \left \langle \tilde{f}_k^{s_k}\right \rangle_{\lambda}\right |^2\right )^{\frac{1}{2}}.
    \end{split}
\end{equation}
The derivation of the third line employs the Cauchy-Schwarz inequality $\langle ab\rangle^2 \leqslant  \langle a\rangle \langle b\rangle$, where \(a = \sqrt{p(\lambda)}\) and $b = \sqrt{p(\lambda)}\textstyle\prod_{k=1}^{N}|\langle \tilde{f}_k^{s_k}\rangle_{\lambda} |$.

For the untrusted parties, the expectation value $\langle \tilde{x}_k \rangle_\lambda$ is determined by 
\begin{equation}
    \begin{split}
        \langle \tilde{x}_k \rangle_\lambda = \displaystyle\sum_{\tilde{x}_k}\tilde{x}_k p(\tilde{x}_k|\tilde{X}_k,\lambda).
    \end{split}
\end{equation}
Given the non-negativity of variances, the untrusted parties fulfill
\begin{equation}
    \begin{split}
        \left |  \left \langle \tilde{f}_k^{\pm } \right \rangle_\lambda\right |^2 &= \left \langle \tilde{x}_k\right \rangle_\lambda^2 + \left \langle \tilde{y}_k\right \rangle_\lambda^2 \leqslant  \left \langle \tilde{x}_k^2\right \rangle_\lambda + \left \langle \tilde{y}_k^2\right \rangle_\lambda,
    \end{split}
\end{equation}
which holds for any hidden variable $\lambda$.
However, for the trusted parties, the expectation value takes the form 
\begin{equation}
    \begin{split}
        \left \langle \tilde{x}_k\right \rangle_{\lambda} =\displaystyle\sum_{\tilde{x}_k}\tilde{x}_k p_Q( \tilde{x}_k | \tilde{X}_k, \rho_{k,\lambda}) = \displaystyle\sum_{\tilde{x}_k}\tilde{x}_k \mathrm{Tr}\left ( E_{\tilde{x}_k | \tilde{X}_k}\rho_{k,\lambda}\right ) = \mathrm{Tr}\left ( \tilde{X}_k\rho_{k,\lambda}\right )=\left \langle \tilde{X}_k\right \rangle_{\lambda},
    \end{split}
\end{equation}
where $E_{\tilde{x}_k | \tilde{X}_k}$ represents the positive operator-valued measure (POVM).
This expression indicates that, under the quantum mechanical measurement framework, the expectation value of the outcomes can be obtained by calculating the expectation of the corresponding observables.
The expectation value $\langle \tilde{y}_k \rangle_\lambda$ satisfies the same relationships.
To establish optimal bounds for the inequalities, we select $X^\perp = Y$ and $Y^\perp = X$.
Additionally, we set $q_k = \min(q_{k,x}, q_{k,y})$, where $q_k \in [1-2\epsilon_k, 1]$.
Under these conditions, the trusted parties satisfy
\begin{equation}
    \begin{split}
        \left |  \left \langle \tilde{f}_k^{\pm } \right \rangle_\lambda\right |^2 &= \left \langle \tilde{X}_k\right \rangle_\lambda^2 + \left \langle \tilde{Y}_k\right \rangle_\lambda^2 = \left \langle X_k\right \rangle_\lambda^2 + \left \langle Y_k\right \rangle_\lambda^2 + 4q_k\sqrt{1-q_k^2}\left \langle X_k\right \rangle_\lambda \left \langle Y_k\right \rangle_\lambda \leqslant  \left ( 1 + 2q_k\sqrt{1-q_k^2}\right )\left ( \left \langle X_k^2\right \rangle_{\lambda} + \left \langle Y_k^2\right \rangle_{\lambda} - C_k\right ) \\ &= \left ( 1 + 2q_k\sqrt{1-q_k^2}\right )\left ( \left \langle x_k^2\right \rangle_{\lambda} + \left \langle y_k^2\right \rangle_{\lambda} - C_k\right ),
    \end{split}
\end{equation}
where the third line utilizes the quantum uncertainty relation of the form $\Delta_{\lambda}^2 \left ( x_k\right ) + \Delta_{\lambda}^2 \left ( y_k\right )\geqslant C_k$, along with the inequality $\left \langle X_k\right \rangle_{\lambda}\left \langle Y_k\right \rangle_{\lambda} \leqslant  \frac{1}{2}\left ( \left \langle X_k\right \rangle_{\lambda}^2+ \left \langle Y_k\right \rangle_{\lambda}^2\right )$.
Here, $C_k$ is a constant that depends on the observables associated with $X_k$ and $Y_k$.

Therefore, we derive the multipartite inequality in the presence of imprecise measurements, expressed in the following form:
\begin{equation}
    \begin{split}
        \left | \left \langle \displaystyle\prod_{k=1}^{N}\tilde{f}_k^{s_k}\right \rangle\right |^2 &\leqslant  \int d\lambda p(\lambda) \displaystyle\prod_{k=1}^{N}\left | \left \langle \tilde{f}_k^{s_k}\right \rangle_{\lambda}\right |^2 \leqslant  \Bigg \langle \displaystyle\prod_{k=1}^{T}\left ( 1+2q_k\sqrt{1-q_k^2}\right )\left ( x_k^2+y_k^2-C_k\right ) \displaystyle\prod_{k=T+1}^{N}\left ( \tilde{x}_k^2+\tilde{y}_k^2\right ) \Bigg \rangle \\ &= \Bigg \langle \displaystyle\prod_{k=1}^{T}\left ( 1+2q_k\sqrt{1-q_k^2}\right )\left ( x_k^2+y_k^2-C_k\right ) \displaystyle\prod_{k=T+1}^{N}\left ( x_k^2+y_k^2\right ) \Bigg \rangle. 
    \end{split}
\end{equation}
For simplicity, we select the two-qubit observables $X_k$ and $Y_k$ to correspond to the Pauli matrices $\sigma_k^x$ and $\sigma_k^y$, respectively.
The constant $C_k$ in the quantum uncertainty relation is consequently determined to be $C_k = 1$.
Based on these relations, we obtain the inequality
\begin{equation}
    \begin{split}
        &\left | \left \langle \displaystyle\prod_{k=1}^{N}\tilde{f}_k^{s_k}\right \rangle\right |^2 \leqslant  2^{N-T} \displaystyle\prod_{k=1}^{T} \left ( 1 + 2q_k \sqrt{1-q_k^2}\right ) \leqslant   2^{N-T}\displaystyle\prod_{k=1}^{T}\left ( 1 + 4\sqrt{\epsilon_k\left ( 1-\epsilon_k\right )} - 8 \epsilon_k\sqrt{\epsilon_k\left ( 1-\epsilon_k\right )}\right ),
    \end{split}
\end{equation}
where $\epsilon_k \in [0, \frac{2-\sqrt{2}}{4}]$.

\end{widetext}

\nocite{*}

\bibliography{BibliSteering}

\end{document}